# Can non-orthogonal bases form stable skyrmionic beams?

Zan Zhang,[1] Zhujun Ye,[2,3] Zhenyu Guo,[2,*] Zihan Liu,[1] Zhuoyue Sun,[1] Zijing Zhang,[1] Jiantao Ma,[4] Katsuya Inoue,[3] Yijie Shen,[2,5,*] Dongmei Deng[1,*]

[1] *Guangdong Provincial Key Laboratory of Nanophotonic Functional Materials and Devices, Guangdong Basic Research Center of Excellence for Structure and Fundamental Interactions of Matter, School of Optoelectronic Science and Engineering, South China Normal University, Guangzhou 510006, China*

[2] *Centre for Disruptive Photonic Technologies, School of Physical and Mathematical Sciences & The Photonics Institute, Nanyang Technological University, Singapore 637371, Singapore*

[3] *International Institute for Sustainability with Knotted Chiral Meta Matter (WPI-SKCM²), Hiroshima University, 1-3-1 Kagamiyama, Higashi-Hiroshima, Hiroshima 739-8526, Japan*

[4] *State Key Laboratory of Optoelectronic Materials and Technologies, School of Physics, School of Electronics and Information Technology, Sun Yat-Sen University, Guangzhou 510275, China*

[5] *School of Electrical and Electronic Engineering, Nanyang Technological University, Singapore 639798, Singapore*

*E-mail: zhenyu.guo@ntu.edu.sg (Z. Guo), yijie.shen@ntu.edu.sg (Y. Shen) and dmdeng@263.net (D. Deng)

**Abstract:** Skyrmions, topologically stable spin textures, have recently garnered significant attention in optics promising robust high-density information transition and nontrivial light-matter interaction. It was believed that the optical skyrmionic beams should be constructed by superposition of two orthogonal spatial modes with orthogonal polarizations to obtain topologically stable propagation. Here, we surprisingly find that propagation-stable skyrmionic beams can still be formed by superpositions of neither orthogonal spatial modes nor orthogonal polarizations. We theoretically present the mechanism to control the stable skyrmionics beams through the hybrid superposition of modes from the Hermite-Gaussian and Laguerre-Gaussian families and experimentally control the longitudinal on-demand dynamics of the skyrmions. This work redefines the topological stability of optical skyrmions, breaks limits and reduces the requirement for manipulating topologically structured light for practical multidimensional implementation of topologically robust information technologies.

## Introduction

Skyrmions are topologically stable quasiparticles characterized by nontrivial textures. Initially proposed to describe the topological structure of nucleons in high-energy and condensed matter physics [1-4], they have subsequently emerged in diverse physical systems [5-10]. Recently, skyrmions have been successfully extended to the optical domain [11-19]. These particle-like topological structures have opened a new frontier in modern optics and significantly expanded the scope of the skyrmion family [14-16]. To date, optical skyrmions have been successfully constructed across various vectorial fields, including spin angular momentum [12,13], electric or magnetic field vectors [11,20], Stokes vectors [21-23], Poynting vectors [24,25], and pseudospin vectors [26,27]. These optical quasiparticles, endowed with unique properties, hold promising applications in fields such as optical communications [28,29], photonic computing [30,31], and super-resolution imaging [12].

Among optical skyrmions, Stokes skyrmions are particularly attractive as information carriers because they can propagate in free space and exhibit topological stability in complex media [32-37]. Their textures arise from the non-separable combination of polarization and spatial modes in the paraxial regime [21-23]. Specifically, the selection of the polarization basis dictates the beam's topology, such as skyrmions and bimerons [22]. The selection of spatial beam modes dictates the spatial morphology and dynamical evolution characteristics of topological quasiparticles, enabling the tailoring of topological light fields with specific functionalities. Various complex topological textures, including skyrmionium [22,38], ring-shaped skyrmions [39,40], and skyrmion lattices [41], have been realized alongside versatile propagation properties such as non-diffraction [42-44], self-healing [42], self-acceleration [41], and propagation invariance [45,46]. However, optical skyrmion beams have been considered to require the superposition of orthogonal basis states to achieve topologically stable propagation [15,16,21-23]. In particular, existing skyrmionic beams have been constructed through the superposition of two spatial modes from the same mode family, such as Laguerre-Gaussian or Bessel-Gaussian modes [15,21-23]. Therefore, whether propagation-stable topological textures can still be formed when these constraints are relaxed remains a key open question. Moreover, current approaches to controlling the evolution of topological textures often depend on complex or predefined optical schemes, such as topological reconfiguration [45] or metasurfaces [46], which constrain their scalability. This highlights the need for alternative approaches that enable more flexible and scalable schemes.

In this work, we propose a mechanism for the stable generation and control of optical skyrmionic beams through the hybrid superposition of modes from the Hermite-Gaussian (HG) and Laguerre-Gaussian (LG) families (Fig. 1). This superposition is not arbitrary, and we provide a detailed explanation of the construction principles necessary to achieve topologically stable textures in hybrid modes. Most importantly, the theoretical and experimental results demonstrate that propagation-stable skyrmionic beams can still be formed even when neither the spatial modes nor the polarization states are orthogonal. The hybrid-mode scheme introduces new physical degrees of freedom, enabling on-demand control of the longitudinal evolution of skyrmionic textures. Specifically, it supports tailored evolution rates and propagation-invariant topologies, with the invariance sustained over infinite distances. This work not only redefines the topological stability of optical skyrmions and reduces the requirements for manipulating topological light, but also paves the way for their applications in robust optical communications and high-capacity encoding.

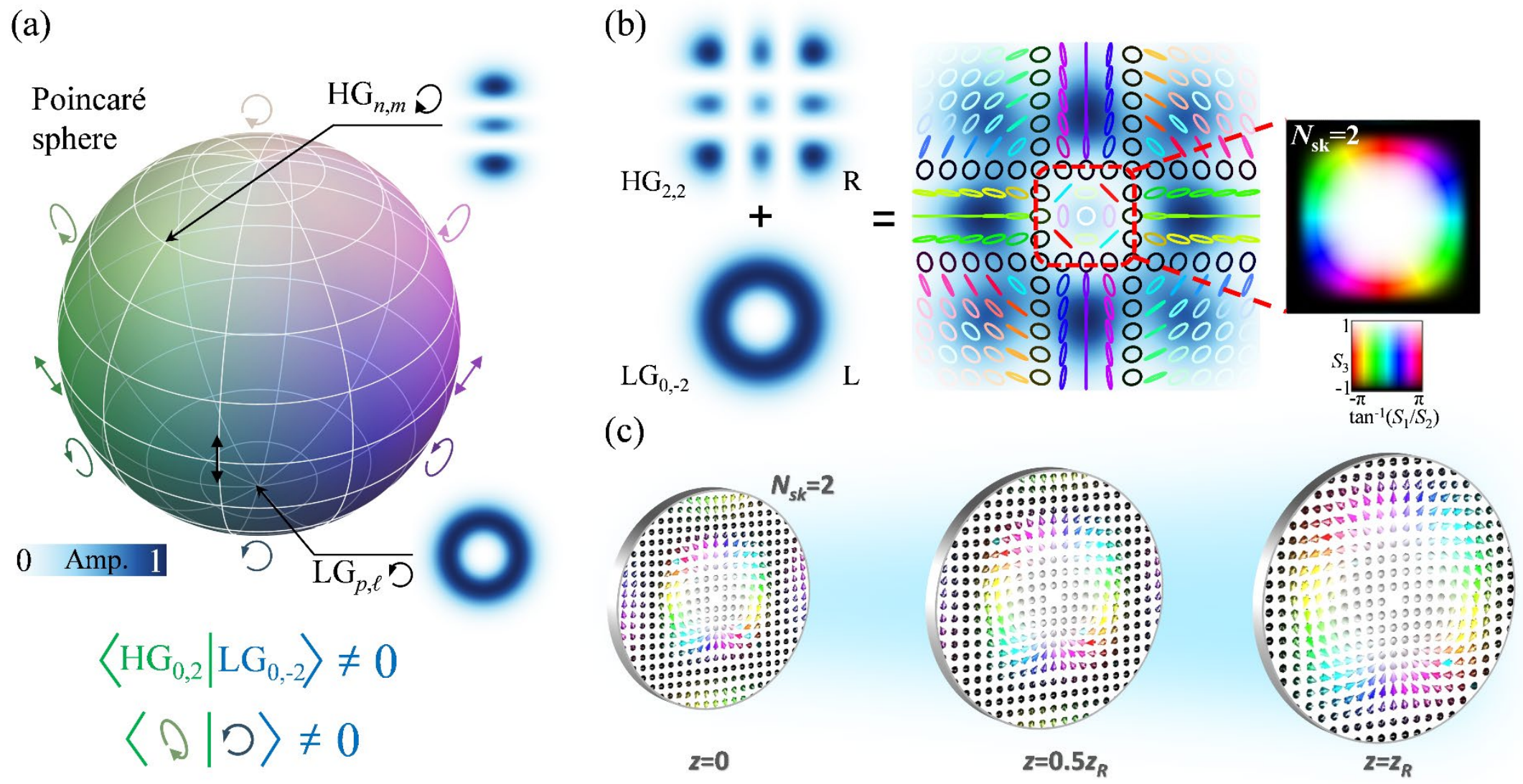

Fig. 1. Schematic illustration of optical skyrmion generation via hybrid-mode superposition. (a) Poincaré sphere in hybrid vector mode (non-orthogonal bases). (b) Generation of a vector mode through the superposition of Hermite-Gaussian ($HG_{2,2}$) and Laguerre-Gaussian ($LG_{0,-2}$) modes under orthogonal circular polarizations, manifesting a nontrivial topological texture with a skyrmion number of $N_{sk}$=2. (c) Realization of Stokes skyrmions exhibiting propagation-invariant topology via tailored hybrid vector modes.

## Results

## Concepts and principle

Stokes skyrmions are typically constructed as full Poincaré vector beams formed by the superposition of distinct spatial modes with orthogonal polarization states, and can be expressed as [21-23]:

$$|\psi\rangle = u_1(\boldsymbol{r})\boldsymbol{e}_1 + e^{i\zeta}u_2(\boldsymbol{r})\boldsymbol{e}_2, \tag{1}$$

where $u_1(\boldsymbol{r})$ and $u_2(\boldsymbol{r})$ denote the complex amplitudes of the spatial modes corresponding to the Jones vectors $\boldsymbol{e}_1$ and $\boldsymbol{e}_2$, respectively, and $\zeta$ represents the global phase difference between these two spatial modes. Specifically, when the Jones vectors $\boldsymbol{e}_1$ and $\boldsymbol{e}_2$ are chosen as orthogonal circular (left- and right-handed) or linear (horizontal and vertical) polarization bases, skyrmionic or bimeronic beams can be realized accordingly. The topological properties of skyrmions can be characterized by the skyrmion number $N_{sk}$ [14-16]:

$$N_{sk} = \frac{1}{4\pi}\iint_{\sigma}\boldsymbol{s}\cdot\left(\frac{\partial \boldsymbol{s}}{\partial x}\times\frac{\partial \boldsymbol{s}}{\partial y}\right)dxdy. \tag{2}$$

where $\boldsymbol{s}$ denotes the normalized Stokes vector, representing the polarization state at each point in the transverse plane, and $\sigma$ represents the region confining the skyrmion. The skyrmion number, which represents the number of times the vector $\boldsymbol{s}$ wraps around the unit parameter sphere, i.e., the Poincaré sphere, can be decomposed into the polarity $p$ and vorticity $m$. More generally, the Jones vectors $\boldsymbol{e}_1$ and $\boldsymbol{e}_2$ do not have to be orthogonal [33], because such polarization orthogonality is not a necessary condition for the formation of optical skyrmions, see Fig. 1(a). The necessary condition for forming Stokes skyrmions is a nontrivial mapping from the transverse plane to the Poincaré sphere, together with the resulting topological wrapping number $N_{\mathrm{sk}}$.

Here, we construct topological textures using hybrid vector modes formed by the coherent superposition of Hermite–Gaussian (HG) and Laguerre–Gaussian (LG) modes [47], as illustrated in Fig. 1(b). We construct optical skyrmions of the form

$$\left|\psi_{n,m}^{p,\ell}\right\rangle = HG_{n,m}\boldsymbol{e}_R + e^{i\zeta} LG_{p,\ell}\boldsymbol{e}_L, \tag{3}$$

where $HG_{n,m}$ and $LG_{p,\ell}$ are the Hermite- and Laguerre-Gaussian modes, respectively. $n$ and $m$ are the indices of the HG mode, $p$ and $\ell$ are the radial and azimuthal indices of the LG mode, and $\boldsymbol{e}_{\mathrm{R(L)}}$ represent the right (left) circular polarization unit vectors, respectively. $\Psi_{LG}^{Gouy} = (2p+|\ell|+1)\tan^{-1}(z/z_R)$ and $\Psi_{HG}^{Gouy} = (n+m+1)\tan^{-1}(z/z_R)$ represent the Gouy phases associated with the propagation of the LG and HG modes, respectively, with $z_{\mathrm{R}}$ being the Rayleigh length of the beam. $M_{\mathrm{L}} = 2p+|\ell|$ and $M_{\mathrm{H}} = n+m$ are defined as the mode orders of the LG and HG modes.

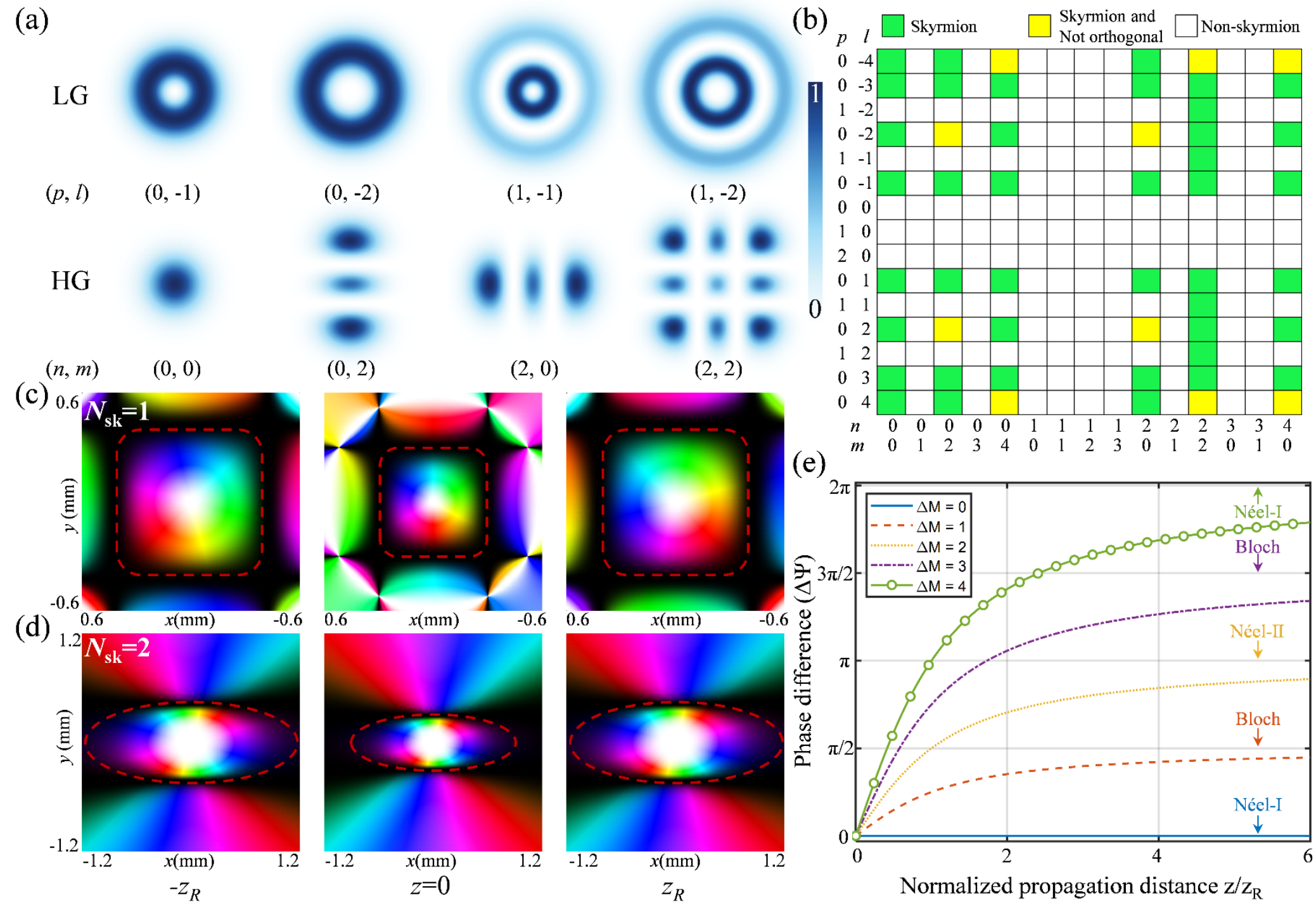

Fig. 2. Theoretical analysis of topological texture generation via hybrid-mode superpositions. (a) Intensity profiles of representative scalar LG and HG modes satisfying the criteria for topological texture generation; (b) For $LG_{p,\ell}$ and $HG_{n,m}$ modes with $M_L$, $M_H \leq 4$, there are 225 possible combinations. Among these, 52 pairs satisfy the topological texture generation conditions, comprising 42 mutually orthogonal pairs (green) and 10 non-orthogonal pairs (yellow); Topological textures of hybrid vector modes at selected propagation distances: (c) $HG_{2,2}\boldsymbol{e}_R+LG_{1,-1}\boldsymbol{e}_L$ and (d) $HG_{0,2}\boldsymbol{e}_R+LG_{0,-2}\boldsymbol{e}_L$; (e) On-demand control over the longitudinal evolution rate of skyrmionic fields by tailoring the mode order difference $\Delta M$ of the hybrid vector modes.

It must be emphasized that not all arbitrary superpositions of HG and LG modes can generate topological textures. First, we require that the HG mode indices $n$ and $m$ both be even integers, ensuring non-zero beam intensity at $r = 0$ (e.g., $HG_{2,0}$, $HG_{2,2}$, and $HG_{4,4}$), as illustrated in Fig. 2(a). The local normalized-state ratio $q$ can then be written as [21]:

$$q = u_{LG}\left(x, y\right)/u_{HG}\left(x, y\right) = \frac{r^{|\ell|}e^{i\ell\phi}L_p^{|\ell|}\left(r^2\right)}{H_{2n}(x)H_{2m}(y)}. \tag{4}$$

The finite skyrmion boundary is defined by the region where the denominator is zero: $H_{2n}(x)H_{2m}(y) = 0$, for which $|q| \to \infty$. On this boundary, $\boldsymbol{s}|_{\partial\sigma} \approx (0,0,-1)$, satisfy the boundary conditions for the skyrmions[32]. Thus, the LG azimuthal index determines the skyrmion winding number, while the HG mode determines the geometry and finite-domain boundary of the skyrmion (see Supplementary Information for details). Based on this and without loss of generality, we confine our analysis to a subset of modes where $M_L$, $M_H \le 4$, including 15 HG modes and 15 LG modes. This leads to 225 possible combinations of modes, 52 of which satisfy the conditions for generating topological textures (highlighted in green and yellow), as shown in Fig. 2(b). These hybrid vector modes can give rise to topological textures with novel geometries, including 'stripe' and 'square' structures, while remaining stable during propagation, as illustrated in Figs. 2(c) and (d).

A detailed analysis of these 52 combinations reveals that 42 (indicated in green) consist of mutually orthogonal modes, analogous to the construction of conventional skyrmionic beams. However, we focus on the remaining 10 combinations (yellow), which involve non-orthogonal superpositions. Because we can express the LG mode as a linear combination of HG modes [48]:

$$LG_{p,\ell}(\rho,\phi,z)=\sum_{k=0}^{N} i^k b(n,m,k) HG_{N-k,k}(x,y,z), \tag{5}$$

where $b(n,m,k)=\sqrt{\frac{(M_*-k)!k!}{2^{N_*}n!m!}}\frac{1}{k!}\left[\frac{d^k(1-t)^n(1+t)^m}{dt^k}\right]_{t=0}$. Therefore, any $LG_{p,\ell}$ is not orthogonal to the $HG_{n,m}$ that appear in its Hermite-Gaussian linear decomposition. Crucially, we find that vector modes constructed from the non-orthogonal superposition can still generate stable topological textures, as shown in Figs. 2(b) and 2(d). This observation suggests that mode orthogonality is not a strict prerequisite for the generation or stable propagation of this new family of hybrid-mode topological textures. To the best of our knowledge, this is the first report to demonstrate that spatial mode orthogonality is not a strict requirement for forming stable topological textures. In the Experimental results section, we further examine the topological stability of spatially non-orthogonal modes when the polarization states are also non-orthogonal.

Next, we theoretically examine the propagation dynamics of topological textures generated by hybrid vector modes. Their evolution is primarily driven by variations in the relative Gouy phase between the constituent modes, expressed as:

$$\Delta\Psi=\Delta M\tan^{-1}(z/z_R), \tag{6}$$

where $\Delta M = M_H - M_L$ denotes the mode order difference between the HG and LG

components. In conventional skyrmionic beams constructed from a single modal family, the constituent modes must possess distinct azimuthal indices $\ell$ to satisfy the skyrmionic mapping requirements. Consequently, these modes typically possess different mode orders, leading to different phase accumulation rates through free space [21]. As a result, the topological textures inherently undergo structural transformations during propagation. Since the spatial indices $n$ and $m$ are independent of the orbital angular momentum carried by the LG mode, specific combinations can be selected to precisely tune the propagation-dependent phase shift. This capability enables on-demand control over the longitudinal dynamic evolution of optical skyrmions.

By tailoring $\Delta M$ through the selection of indices $n$ and $m$, we achieve on-demand control over the longitudinal evolution rates. As illustrated in Fig. 2(e), this approach allows for customized transitions between diverse topological textures, such as the transition from Néel-I-type to Bloch-type (for $\Delta M$=1 and 3), Néel-II-type ($\Delta M$=2), and Néel-I-type ($\Delta M$=4) skyrmions. Figure 2(c) demonstrates the spontaneous evolution of topological textures, which transition from a Néel-I type ($z = 0$) to an intermediate-type ($z = z_R$), with Bloch-type skyrmions emerging in the limits $z = \pm\infty$ (not shown), consistent with the orange dashed line in Fig. 2(e). Conversely, by setting $\Delta M$=0, Fig. 2(d) verifies that the skyrmion topology exhibits propagation invariance; the texture undergoes only self-similar spatial stretching, while the skyrmion number remains robust, consistent with the blue solid line in Fig. 2(e). This approach, centered on the modulation of the mode order difference via hybrid modes, is fundamentally distinct from previous methods for manipulating structured vector beams via Bessel modes [45]; this method can maintain topological invariance over infinitely long distances.

## Experimental results

We constructed an experimental setup based on a spatial light modulator (SLM), using an SLM hologram with a complex amplitude modulation scheme in combination with a Sagnac interferometer to generate hybrid vector mode optical skyrmions (refer to the experimental methods section for details). First, by employing different hybrid vector modes, we experimentally generated first- and second-order hybrid-mode skyrmions, including ‘stripe’ and ‘square’ topological textures, among others, as shown in Fig. 3(a). These results confirm the feasibility of generating skyrmionic beams from hybrid modal families.

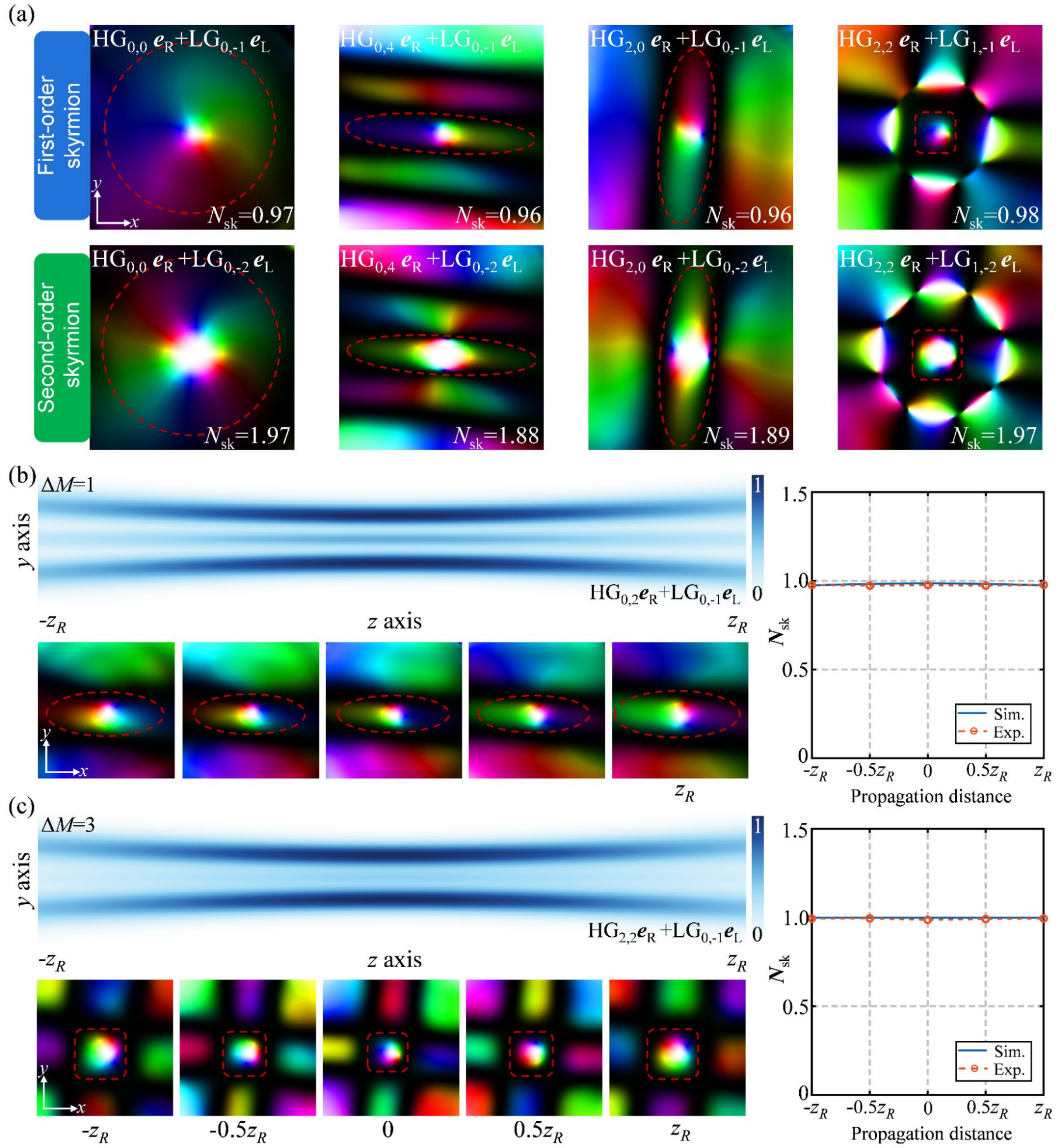

Fig. 3. Experimental and theoretical results of topological textures formed by hybrid vector modes. (a) Experimental topological textures generated by different hybrid vector-mode configurations. The upper-left inset in each texture shows the corresponding vector-mode configuration, and the lower-right value gives the experimentally measured skyrmion number. (b, c) Theoretical and experimental results of topological textures with different propagation-evolution rates: (b) $\Delta M$= 1 and (c) $\Delta M$= 3. In both cases, the upper-left panel shows the simulated normalized intensity distribution in the longitudinal $y$ - $z$ plane, with the corresponding vector mode shown in the lower-right inset. The lower-left panel shows the experimentally observed textures at selected transverse $x$ - $y$ planes. The right panel shows the simulated (line) and experimental (point) skyrmion numbers as a function of propagation distance.

Next, we experimentally demonstrated the evolution of five representative hybrid-mode skyrmionic textures during free-space propagation, with mode-order differences of $\Delta M$ = 0, 1, and 3. This demonstrates on-demand control over the longitudinal evolution of the skyrmion field, including both tunable evolution rates (Fig. 3) and propagation-invariant topologies (Fig. 4). First, Stokes skyrmions with two distinct propagation-evolution rates ($\Delta M$ = 1, 3) were experimentally generated, and their topological textures were measured at various transverse planes along the propagation direction, as shown in Fig. 3. The first example was realized via the superposition of orthogonal circular polarizations using the $HG_{0,2}$ and $LG_{0,-1}$ modes, as defined in Eq. (3). The second example was generated by superimposing the $HG_{2,2}$ and $LG_{0,-1}$ modes. These two examples yield "strip" and "square" topological textures with $N_{sk}$ = 1, as shown in Figs. 3(b) and 3(c), respectively. Owing to the mode-order difference, the skyrmionic beam acquires a relative Gouy phase shift during free-space propagation, which drives the spontaneous evolution of the topological texture while preserving the skyrmion number. These two textures undergo spontaneous evolution accompanied by diffraction-induced spatial stretching. However, the longitudinal evolution rates differ between them; a larger $\Delta M$ leads to a faster rotation of the topological texture. Specifically, from the source (z = 0) to the Rayleigh range ($z = z_R$), the skyrmionic beams accumulate relative Gouy phase shifts of $\pi/4$ for $\Delta M$=1 and $3\pi/4$ for $\Delta M$=3, respectively.

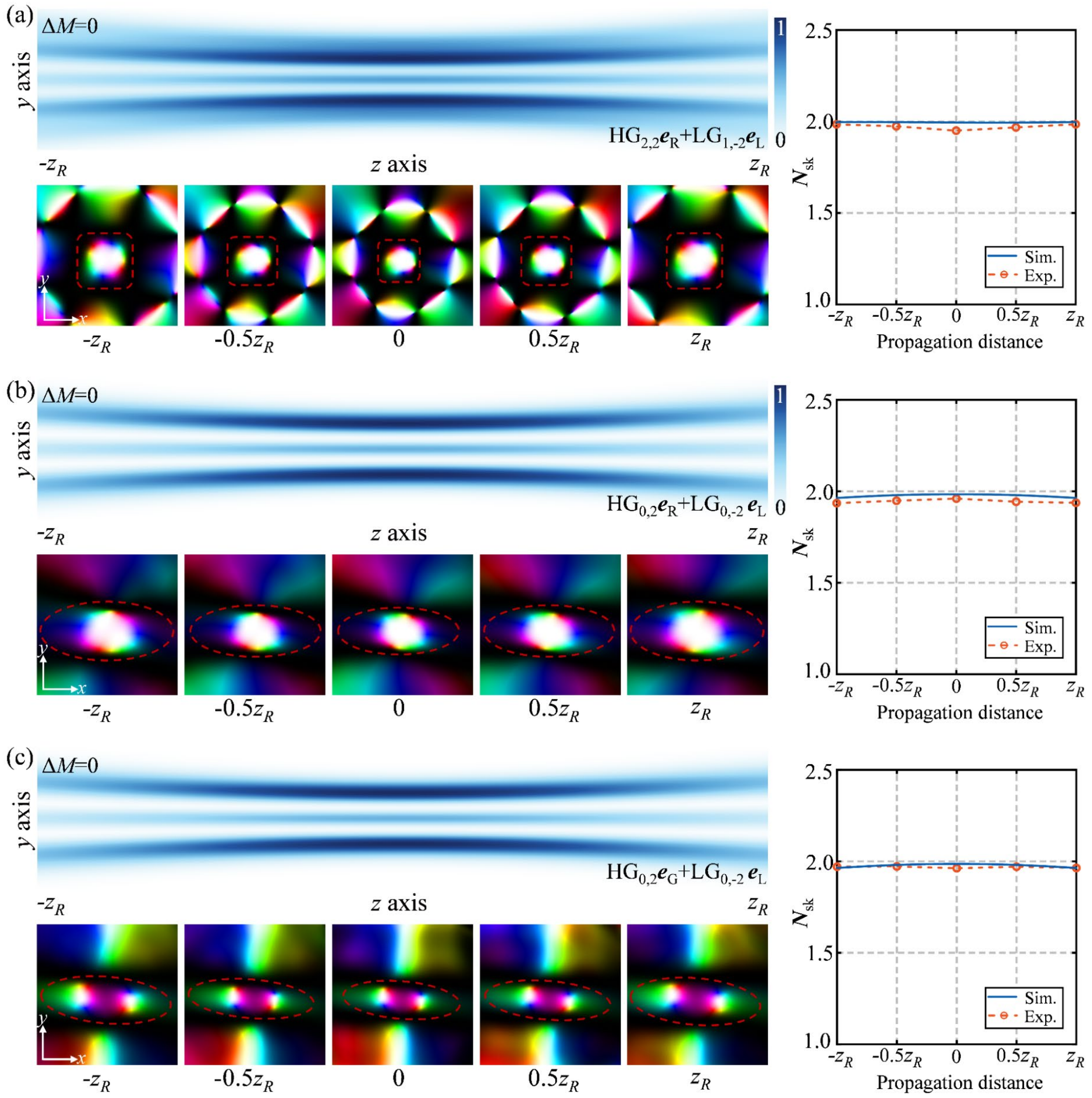

Fig. 4. Theoretical and experimental results for propagating invariant topologies. (a) Both spatial modes and polarization are orthogonal; (b) Spatial modes are non-orthogonal, but polarization is orthogonal; (c) Neither spatial modes nor polarization are orthogonal. Left (top): Simulated normalized intensity distributions in the longitudinal $y$ - $z$ plane, the vector mode in the bottom right. Left (bottom): experimentally observed topological textures at selected transverse $x$ - $y$ planes. Right: variation of simulated (line) and experimental (point) skyrmion numbers as a function of propagation distance. Here, $\boldsymbol{e}_{\mathrm{G}} = \cos(\gamma)\, \boldsymbol{e}_{\mathrm{R}} + \sin(\gamma)\, \boldsymbol{e}_{\mathrm{L}}$, and $\gamma = \pi/6$.

Then, by selecting hybrid vector modes with identical mode orders, we experimentally demonstrated propagation-invariant topological textures. Specifically, we constructed three representative hybrid-mode topological textures: the first was formed by superposing orthogonal spatial modes with orthogonal polarization states [Fig. 4(a)], the second by superposing non-orthogonal spatial modes with orthogonal polarization states [Fig. 4(b)], and the third by superposing non-orthogonal spatial modes with non-orthogonal polarization states [Fig. 4(c)]. In the experiments, we used vectorial superpositions of the $HG_{2,2}$ and $LG_{1,-2}$ modes (orthogonal spatial modes), as well as the $HG_{0,2}$ and $LG_{0,-2}$ modes (non-orthogonal spatial modes), and tuned the polarization orthogonality of the non-orthogonal spatial modes, thereby realizing second-order skyrmion textures [Figs. 4(a) and 4(b)] and a biskyrmion texture [Fig. 4(c)]. We measured their topological textures at different transverse planes along the propagation direction, as shown in Fig. 4.

In all three cases, because the constituent modes had the same mode order, no relative Gouy phase difference accumulated between them during propagation. As a result, the skyrmionic textures remained propagation-invariant, undergoing only diffraction-induced spatial stretching, while the skyrmion number remained highly stable throughout propagation. More importantly, these results indicate that propagation-stable skyrmionic beams can still be formed without requiring either orthogonal spatial modes or orthogonal polarization states, and even when both the spatial modes and polarization states are non-orthogonal. This is surprising because most previously reported free-space Stokes skyrmion schemes rely on orthogonal bases. Our work overcomes these previous limitations and lowers the barrier to practical multidimensional applications.

Finally, to quantify the robustness of the hybrid-mode topological textures against polarization non-orthogonality, we varied the polarization state of the HG mode while keeping that of the LG mode fixed: $|\psi\rangle = HG_{n,m}\boldsymbol{e}_G + LG_{p,\ell}\boldsymbol{e}_L$, where $\boldsymbol{e}_G = \cos(\gamma)\, \boldsymbol{e}_R + \sin(\gamma)\, \boldsymbol{e}_L$ and $\gamma \in [0, \pi/2]$ controls the transition from orthogonal polarization states at $\gamma=0$ to fully parallel polarization states at $\gamma=\pi/2$. We investigated the robustness of two hybrid-mode topological textures against polarization non-orthogonality: $HG_{0,0}\, \boldsymbol{e}_G + LG_{0,-2}\, \boldsymbol{e}_L$ and $HG_{0,2}\, \boldsymbol{e}_G + LG_{0,-2}\, \mathbf{e}_L$. We first analyzed the evolution of the topological textures as a function of $\gamma$, as shown in Fig. 5(a). As the polarization states changed from orthogonal to non-orthogonal, both topological textures exhibited a transition from second-order skyrmion textures to biskyrmion textures. We then analyzed the variation of $N_{sk}$ with $\gamma$ to quantify the robustness of the topological textures against polarization non-orthogonality; the simulated and experimental results are shown in Fig.

5(b). It can be seen that, during the transition from orthogonal to non-orthogonal polarization states, $N_{\mathrm{sk}}$ was still well maintained for both hybrid-mode configurations, even near the fully non-orthogonal polarization condition at γ=5π/12. When γ =π/2, corresponding to fully parallel polarization states, the skyrmion beam degenerates into a uniformly polarized state, namely a topologically trivial state. The experimental results are in good agreement with the simulations, as shown in Fig. 5(b). These results provide strong evidence of the robustness of hybrid-mode topological textures under non-orthogonal polarization conditions.

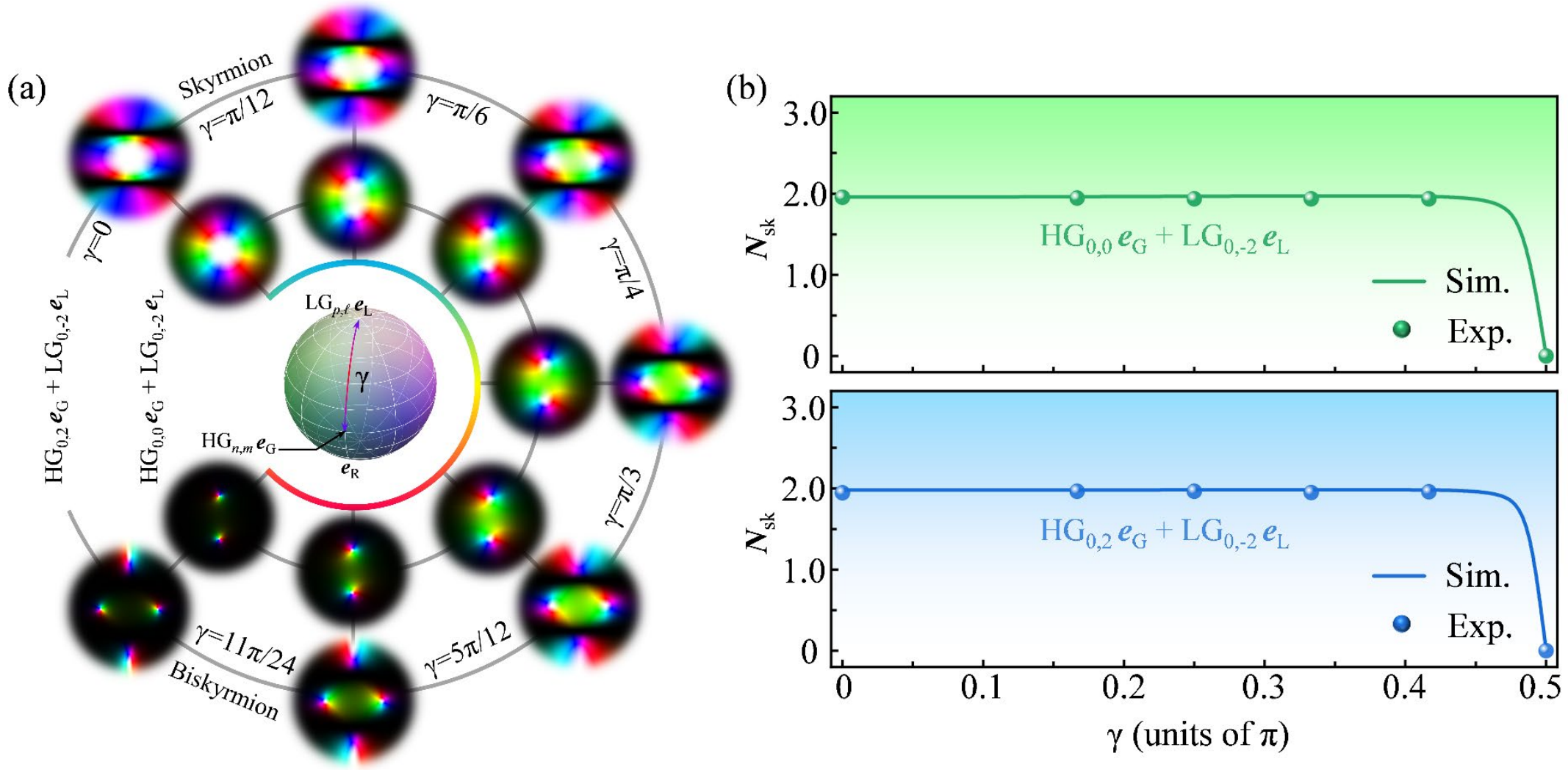

Fig. 5. Robustness analysis of hybrid-mode topological textures under non-orthogonal polarization conditions. (a) Simulation results of topological textures as a function of γ; (b) Simulation (lines) and experimental (point) skyrmion numbers as a function of γ.

## Discussion

In summary, we theoretically present the mechanism to control the stable skyrmionic beams through the hybrid superposition of modes from the Hermite-Gaussian and Laguerre-Gaussian families. Through precise inter-family phase modulation, our approach enables active control over the longitudinal dynamics of the resulting topological textures. Distinct from previously reported free-space skyrmionic beams, an OAM-independent mode index introduces a new physical degree of freedom for precisely tailoring the longitudinal evolution of the topology, allowing controllable evolution rates, including propagation-invariant skyrmionic fields whose topological charge and texture remain unchanged over appreciable free-space propagation distances.

Furthermore, we emphasize that using HG modes allows us to define a compact finite region where the skyrmion number is quantized to an integer, which removes the need to integrate up to infinity.

More importantly, we further show that the formation of propagation-stable skyrmionic beams does not strictly rely on the superposition of two orthogonal spatial modes with orthogonal polarization states. Even under conditions involving non-orthogonal spatial modes or non-orthogonal polarization states, and even when both the spatial modes and polarization states are non-orthogonal, propagation-stable skyrmionic beams can still be formed. This result overcomes the conventional reliance on orthogonal bases in the construction of skyrmionic beams and provides new insight into the topological robustness of optical skyrmions. Finally, using an SLM-based vector beam generator, we experimentally realize the proposed strategy, demonstrating topological textures with on-demand controllable longitudinal dynamics, stable topological textures formed through non-orthogonal bases superpositions, and the robustness of the topological textures under polarization non-orthogonality.

This work holds broad implications on several fronts. First, forming topological textures from modes belonging to different mode families introduces additional degrees of freedom while preserving the topological protection of the textures, thereby offering a route toward robust and high-capacity optical communications [49,50]. Second, our results provide both a theoretical and an experimental foundation for the tunable longitudinal evolution and manipulation of skyrmionic beams, opening promising opportunities for optical sensing [51,52] and establishing programmable foundations for topological structuring in both classical and quantum wave systems [53]. Our scheme can be further extended to explore broader forms of skyrmions, including higher-order beam modes, faster evolution rates, and richer topological structures. Crucially, this work redefines the topological stability of optical skyrmions, overcomes previous limitations and reduces the requirement for manipulating topologically structured light for practical multidimensional implementation of topologically robust information technologies.

## Experimental methods

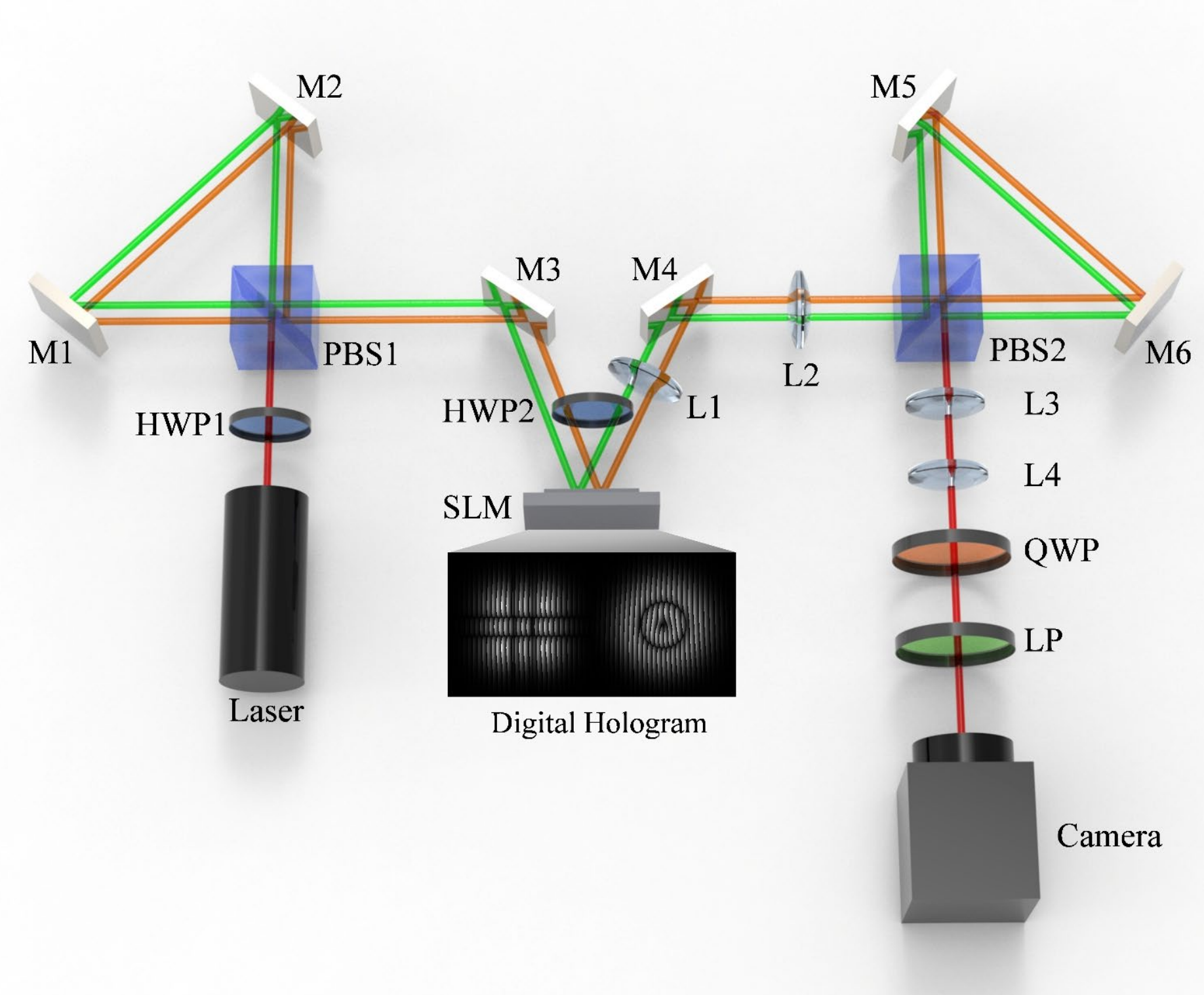

Fig.6. Experimental setup for hybrid-mode topological texture generation. HWP: half-wave plate, PBS: polarizing beam splitter, M: mirror, SLM: spatial light modulator, L: lens, QWP: quarter-wave plate, LP: linear polarizer.

To experimentally generate the desired topological textures of the hybrid vector modes, we employ a Sagnac interferometer with a reflective phase-only spatial light modulator (SLM, PLUTO-2.1-NIR-113) to generate arbitrary polarization textures. The experimental setup is shown in Fig. 6. The first Sagnac interferometer separates the incident linearly polarized beam ($\lambda$ = 780 nm) into two counter-propagating orthogonally polarized components. A half-wave plate aligns the polarization with the SLM modulation axis, while an encoded hologram independently controls the amplitude and phase of each component via complex-amplitude modulation. The first 4f system (L1–L2) acts as a spatial filter. After coherent recombination in the second Sagnac interferometer, the beam is relayed through a second 4f system (L3 and L4), whose output plane defines the initial plane (z = 0). Finally, the optical skyrmions are experimentally reconstructed via Stokes polarimetry using a quarter-wave plate, a linear polarizer, and a CMOS camera.

**Acknowledgements:** This work was supported by the Natural Science Foundation of China (Grant No. 12174122), and the Guangdong Provincial Natural Science Foundation of China (Grant Nos. 2024A1515010556 and 2026A1515011961). Y. Shen acknowledges the support from Singapore Ministry of Education (MOE) AcRF Tier 1 grants (RG157/23 & RT11/23), Singapore Agency for Science, Technology and Research (A*STAR) (M24N7c0080 & R25J4IR110), and Nanyang Assistant Professorship Start Up grant.

**Disclosures:** The authors declare no conflicts of interest.

**Data availability:** The data that support the findings of this study are available from the corresponding author upon reasonable request.